\documentclass[12pt,english]{article}

\usepackage{geometry}
\usepackage[titletoc,title]{appendix}
\usepackage{ragged2e}
\usepackage{seqsplit}
\usepackage{setspace}
\usepackage{braket}
\usepackage{academicons}
\usepackage{xcolor}
\definecolor{orcidlogocol}{HTML}{A6CE39}

\raggedright

\usepackage{sectsty}
\allsectionsfont{\normalsize\raggedright\centering}

\newcommand{\dif}{\mathrm{d}}%

\usepackage[]{tocloft}
\addtocontents{toc}{\cftpagenumbersoff{section}}
\addtocontents{toc}{\cftpagenumbersoff{subsection}}

\makeatletter
\renewcommand{\@cftmaketoctitle}{}
\makeatother

\newlength{\myl}%
\newcommand{\SUM}[2]{{\setlength{\myl}{\widthof{$\displaystyle\sum_{#1}^{#2}$}*\real{0.5}-\widthof{$\displaystyle\sum$}*\real{0.5}}\sum_{#1}^{#2}\;\hspace{-\the\myl}}}
\newcommand{\INT}[3]{\settowidth{\myl}{$\displaystyle\int_{#1}^{#2}$}{\int_{#1}^{#2}\;\;\;\hspace{-\the\myl}\dif #3}\,}
\newcommand{\TINT}[3]{\settowidth{\myl}{$\int_{#1}^{#2}$}{\int_{#1}^{#2}\!\ifthenelse{\equal{#1#2}{}}{}{\;\;\;\;\hspace{-\the\myl}}\dif #3}\,}%
\newcommand{\EINT}[3]{\settowidth{\myl}{$\int_{#1}^{#2}$}{\int_{#1}^{#2}\;\;\;\,\hspace{-\the\myl}\dif #3}\,}

\usepackage[]{footmisc}

\usepackage{natbib}
\usepackage{url}
\makeatletter
\def\url@leostyle{%
    \def\UrlFont{\sf}}{\def\UrlFont{\small\ttfamily}}
\makeatother
\bibpunct
{(} 
{)} 
{,} 
{a} 
{} 
{;} 

\usepackage{authblk}

\usepackage{graphicx}
\usepackage{babel}

\usepackage{amsmath}
\usepackage{amsfonts}

\usepackage{framed}

\usepackage{xifthen}


\usepackage{txfonts}
\usepackage[T1]{fontenc}

\newcommand{\citealtbjps}[2][]{\ifthenelse{\equal{#1}{}}{\citeauthor{#2} [\citeyear{#2}]}{\citeauthor{#2} \citeyear{#2}, #1}}
\newcommand{\citepbjps}[2][]{\ifthenelse{\equal{#1}{}}{(\citeauthor{#2} \citeyear{#2})}{(\citeauthor{#2} \citeyear{#2}, #1)}}
\newcommand{\citeyearbjps}[2][]{\ifthenelse{\equal{#1}{}}{\citeyear{#2}}{\citeyear{#2}, #1}}
\newcommand{\citeyearparbjps}[2][]{\ifthenelse{\equal{#1}{}}{(\citeyear{#2})}{(\citeyear{#2}, #1)}}
\newcommand{\citeposbjps}[2][]{\ifthenelse{\equal{#1}{}}{\citeauthor{#2}'s (\citeyear{#2})}{\citeauthor{#2}'s (\citeyear{#2}, #1)}}

\usepackage{hyperref}
\usepackage[capitalize,noabbrev]{cleveref}

\newcommand{\ZT}[1]{`#1'}%
\newcommand{\Nabla}{\vec{\nabla}}%

\usepackage[normalem]{ulem}
\usepackage{color}

\newcommand{\orcid}[1]{
  \href{https://orcid.org/#1}{\textcolor{orcidlogocol}{\aiOrcid}}
}

\begin{document}

\title{How to distinguish between indistinguishable particles}
\author{}
\author{Michael te Vrugt}
\date{Forthcoming in \textit{The British Journal for the Philosophy of Science}}
\maketitle
\vspace*{-13.5mm}

\thispagestyle{empty}
\justifying
\begin{abstract}
A long and intense debate in philosophy is concerned with the question whether there can be haecceistic differences between possible worlds, that is, nonqualitative differences that only arise from different de re representations. According to haecceitism, it can give rise to a different situation if the positions of two qualitatively identical particles are exchanged, while according to anti-haecceitism, this is not the case. It has been suggested that classical statistical mechanics might provide evidence for one of these positions. However, most philosophers of physics argue that it does not. In this article, we show that order-preserving dynamics, a novel method from statistical mechanics developed for the description of nonergodic systems, changes this situation: It is intrinsically haecceistic and makes different experimental predictions than non-haecceistic alternatives. Thereby, it provides an empirical argument for the existence of modality de re.\\
\textbf{ORCID:} https://orcid.org/0000-0002-1139-3925
\end{abstract}

\mbox{} \\

\tableofcontents

\mbox{} \\

\section{\label{intro}Introduction}
Consider a system of two qualitatively identical particles $A$ and $B$. Does it make any difference if their positions are exchanged? Despite the simplicity of this problem, questions of this form have attracted a great deal of philosophical attention since medieval times \citep{Cross2014}. Roughly speaking, there are two possible answers: According to haecceitism, it can make a difference, since after the exchange each particle is at a different position. This requires that two qualitatively identical situations can differ in what they represent de re about a certain particle. According to anti-haecceitism, the fact that these two situations are qualitatively identical implies that there can be no difference between them. This is also implied by a suitable version of Leibniz' principle of the identity of indiscernibles \citep[p. 224]{Lewis1986}.

In modern philosophy, it has been suggested that classical statistical mechanics might provide provide an argument for haecceitism. The reason is that, according to what \citet{FrenchR2003} call the \ZT{received view}, it is assumed in Maxwell-Boltzmann statistics that exchanging two particles leads to a different arrangement - in contrast to quantum statistics, where this exchange makes no difference. However, \citet{Huggett1999} claims that since classical mechanics can be formulated in two different frameworks, one of which is haecceistic and one of which is anti-haecceistic (see \cref{classical}), classical statistical physics is neutral regarding the problem of haecceitism. This also implies that quantum mechanics is not innovative in this regard.

While previous philosophical approaches to this issue solely consider equilibrium statistical mechanics, the field-theoretical approach to statistical mechanics develops its full power in the more general case of nonequilibrium systems. A paradigmatic example of a nonequilibrium statistical field theory is dynamical density functional theory (DDFT) \citep{teVrugtLW2020}. Recently, \citet{WittmannLB2020} have developed a variant of DDFT known as order-preserving dynamics (OPD). OPD for one-dimensional systems is characterized by a statistical average in which of the two possible configurations \ZT{particle $A$ is on the left of particle $B$} and \ZT{particle $B$ is on the left of particle $A$} only one is allowed, thereby explicitly treating them as different. This, they show, is required for accurate empirical predictions.

In this article, we show that OPD allows us to develop an empirical argument for haecceitism. The reason is that, in OPD, particle labels are used as rigid designators, and situations that only differ in their de re representation are given different statistical weights. Since OPD is designed for (and required for) the accurate description of certain experiments in a field-theoretical framework, it therefore provides inductive evidence for the existence of modality de re.

This article is structured as follows: In \cref{haecc1}, the problem of haecceitism is introduced. The literature on the relation between haecceitism and statistical mechanics is discussed in \cref{classical}. OPD is introduced in \cref{opd}. In \cref{haecc2}, we then construct an inductive argument for haecceitism and de re modality based on OPD. The results are connected to the debate around the Gibbs paradox in \cref{gibbs}. Possible implications of quantum indistinguishability are discussed in \cref{qm}. We conclude in \cref{conclusion}.

\section{\label{haecc1}Haecceitism and rigid designation}
The metaphysical problem we wish to discuss in this work is that of haecceitism. We here define it following \citet[p. 221]{Lewis1986} (also cited in \citet[p. 7]{Huggett1999}):

\small{If two worlds differ in what they represent de re concerning some individual, but do not differ qualitatively in any way, I shall call that a haecceistic difference. Haecceitism, as I propose to use the word, is the doctrine that there are at least some cases of haecceitistic difference between worlds. Anti-haecceitism is the doctrine that there are none. \citep[p. 221 (emphasis removed)]{Lewis1986}}

Discussions of haecceitism have a long tradition in philosophy (see \citet{Cross2014,Saunders2013}). It can be defined in various ways \citep[p. 407-408]{Gordon2002}. \citet{Adams1979} understands it in terms of a \ZT{primitive thisness} (haecceity) that is attributed to objects, whereas \citet{Lewis1986} understands it in terms of the definition cited above. \citet[p. 409]{Gordon2002} has argued that haecceitism in Lewis' sense implies the existence of primitive thisness since haecceitism without haecceities is difficult to make sense of. This view will be adopted here, see \citet[p. 225]{Lewis1986} for an alternative position.

A further topic that is relevant here is the theory of (rigid) designation. It became popular through the work of \citet{Kripke1980}, who explained it using the famous example of a person pointing at Richard Nixon and saying \ZT{That's the guy who might have lost} \citep[p. 41]{Kripke1980}. This is a statement de re about Nixon, according to which there are possible worlds in which he lost the presidential election. \ZT{Nixon}, which is the name of this individual, is a rigid designator, designating Nixon in all possible worlds \citep[p. 48]{Kripke1980}. In contrast, \ZT{the winner of the election}, which is a definite description, is a non-rigid designator, since there are possible worlds in which someone else won the election. According to Kripke, it is possible to stipulate possible worlds based on de re representation. For example, in a situation in which two dice $A$ and $B$ are thrown \citep[pp. 16-17]{Kripke1980}, the probability of getting five and six is 1/18 since, of all possible thirty-six configurations, there are two in which one die shows five and one six. These differ in which die shows five, that is, the possible worlds differ in what they represent de re about $A$ and $B$.

It thus appears as if Kripke is a typical haecceitist. However, \citet[pp. 222-227]{Lewis1986}, who is both a supporter of Kripkean specification and an anti-haecceist, argues that Kripke is in fact neutral regarding this issue. Anti-haecceists believe that representation de re supervenes on qualitative differences, however, this does not prevent them from ignoring these qualitative differences and specifying possible worlds by what they represent de re. Regarding the dice example, the worlds \ZT{die $A$ five, die $B$ six} and \ZT{die $A$ six, die $B$ five} correspond, for an anti-haecceitist, to what Lewis calls \ZT{miniworlds} \citep[p. 226]{Lewis1986}. In speaking about them, one ignores qualitative differences between the dice such as their origin and position. See also \citet{Maidens1998} for a discussion of this issue in the context of statistical mechanics.

Note, however, that it is essential for the anti-haecceist that there are qualitative differences in order for differences de re to be possible. Moreover, while haecceitism is not identical to the view that one can meaningfully speak about representation de re \citep[p. 222]{Lewis1986}, it does (by definition) imply the possibility of de re representation. In particular, if we have found a haecceistic difference between two possible worlds, we have thereby also shown that modality de re is possible. Moreover, only a haecceist can construct two different possible worlds that only differ in what they represent de re about a certain entity via Kripkean specification.

Since our argumentation relies on Lewis' definition of haecceitism, which is based on the idea of modality de re, it is helpful to briefly discuss what modality de re means for Lewis in particular (which should not be understood as an endorsement of this particular understanding). We follow \citet[pp. 221-222]{BeebeeM2015}. According to \citet{Quine1963}, the idea of modality de re as appearing in modal logic makes no sense since whether something is true about (say) Nixon in a possible world depends on how Nixon is designated. Modality de re, Quine argues, is based on Aristotelian essentialism, which he considers a highly problematic view. \citet{Lewis1968} solves this problem using his \ZT{counterpart theory}: If we say that \ZT{Nixon might have lost the election}, then what we actually mean according to Lewis is \ZT{There is a possible world containing a counterpart of Nixon that has lost the election}. This counterpart relation is not identity \citep[p. 222]{BeebeeM2015}, that is, it is not Nixon himself who lost the election in the possible world, but Nixon's counterpart. Whether a particular individual in a possible world is Nixon's counterpart then depends on similarity relations. These similarity relations, which depend on qualitative resemblance, are not uniquely determined. For example, if we pick out the Nixon of a possible world by similarity of origin, there might be a possible world in which Nixon dies as a child. If one focuses on similarity of career instead, there might be a world where Nixon was born to different parents (here, \ZT{Nixon} can have a totally different origin as long as he has a similar political career). Due to the similarity requirement, everything is its own counterpart. (Discussion adapted from \citet{HallRS2021}.)

To figure out whether situations defined by the configuration of particles in statistical mechanics are identical, it is helpful to consider some criteria that allow to distinguish between physical particles. Following \citet[p. 395]{Gordon2002}, we call particles observationally distinguishable if exchanging them makes a measurable difference, and conceptually distinguishable if it is in principle possible to think of them as being different. As a possible basis for (conceptual) distinguishability, \citet[p. 396]{Gordon2002} mentions (1) properties of the particle, (2) spatiotemporal trajectory or location\footnote{One could legitimately object to this classification that the spatiotemporal location is also a property of the particle. However, the spatiotemporal location has a special status among the particle's properties since, if particles are impenetrable, it is unique to one specific particle.}, and (3) (the possibility of giving) names that only refer to this particular entity. Regarding (1) we can further distinguish between intrinsic properties such as mass and charge and state-dependent properties such as momentum and energy. Intrinsic properties only determine the particle species, and state-dependent properties are not generally sufficient for individuation since it is possible that two particles of the same species also have the same momentum. The only state-dependent property that can be useful for the classical particles considered here is the spatiotemporal trajectory, which, assuming impenetrability (as we shall do for the rest of this article) cannot be identical for two different particles. Naming the particles requires that they possess some form of \ZT{thisness} or individuality\footnote{This can cause difficulties in the description of quantum objects, whose individuality is a matter of debate, using first-order logic, which requires names \citep{Naeger2020}.} \citep[pp. 397-398]{Gordon2002}, whether it is due to qualitative or to primitive thisness.
 
\section{\label{classical}Metaphysics of classical statistical mechanics}
A variety of authors have discussed whether statistical mechanics might provide answers to the problem of haecceitism discussed in \cref{haecc1} (see \citet[pp. 51-64]{FrenchK2006} for an introduction). To see the motivation, consider (following \citet{French2019}) two classical particles $A$ and $B$ that can be distributed over two boxes one and two. How many options are there for doing this? Clearly, there are four: We can put $A$ and $B$ in box one, we can put $A$ and $B$ in box two, we can put $A$ in box one and $B$ in box 2, or we can put $A$ in box two and $B$ in box one. What is important regarding the last two options is that it is assumed here that it makes a difference if $A$ and $B$ are exchanged, that is, \ZT{$A$ in one and $B$ in 2} and \ZT{$A$ in two and $B$ in one} are two different situations. Consequently, there is a probability of 1/2 of having one particle in each box, whereas the other two configurations (both particles on the left and both particles on the right) have a probability of 1/4. This is referred to as \ZT{Maxwell-Boltzmann statistics}. 

These considerations about particle statistics appear to be quite similar to Kripke's dice example, and they indeed are. However, in classical mechanics, the world is fully described once the configuration (position and momenta) of all particles is known \citep{Maidens1998}. Hence, there are no qualitative differences between two worlds in which (other things being equal) two indistinguishable classical particles are exchanged, such that an anti-haecceist cannot consider them to be different. Lewis' rescue of anti-haecceitism based on miniworlds is therefore blocked. The fact that Maxwell-Boltzmann statistics nevertheless appears to require these worlds to be different seems to suggest that classical\footnote{The quantum case is discussed in \cref{qm}.} statistical mechanics requires haecceitism. 

An important line of argument which claims that classical statistical mechanics does not necessarily require haecceitism has been developed by \citet{Huggett1999}. It is based on the fact that there are two different ways of describing the dynamics of classical particles: First, there is phase space, which (following the literature), we refer to as \ZT{$\Gamma$-space}. Here, it is assumed that there are $N$ individual particles that have a certain number of degrees of freedom (usually position $\vec{r}$ and momentum $\vec{p}$). The dynamical equations then describe these degrees of freedom as a function of time. The second option is the so-called \ZT{Z-space}: Here, one characterizes the system's configuration by simply saying how many particles are in which state (formally, one passes to the quotient space). Here, we assume for simplicity that the particles are overdamped, such that the phase space coordinates involve solely the positions of the particles. In this case, the Z-state can be specified using a dichotomic field $\rho(\vec{r})$ that takes the value one if there is a particle at position $\vec{r}$ and zero otherwise. In a continuous space containing $N$ particles, this field is defined as $\sum_{i=1}^{N} \delta(\vec{r}- \vec{r}_i)$ with the Dirac delta distribution $\delta$ and the position vector of the $i$th particle $\vec{r}_i$ \citep{Dean1996}. Importantly, an exchange of two qualitatively indistinguishable particles leads to a new state in $\Gamma$-space, but makes no difference for the Z-state.

The problem of the two particles in two boxes provides an excellent case study for the problem of haecceitism introduced in \cref{haecc1}: Let us assume that the particles $A$ and $B$ are qualitatively identical (regarding properties such as mass, shape, ...). In this case, the situation where $A$ is in box one and $B$ is in box two is qualitatively identical to the situation where they are exchanged. There are now two possible positions regarding the question whether these situations are completely identical:
\begin{enumerate}
    \item Haecceitism: Although these situations are qualitatively identical, they are different: In one case, particle $A$ is in box one, whereas in the other case, it is in box two. Therefore, there is a de re difference between these situations. This is represented by $\Gamma$-space, which describes these situations differently: In one case, the position $\vec{r}_A$ of particle $A$ is in box one (and the position $\vec{r}_B$ of particle $B$ is in box two). In the other case, the position $\vec{r}_A$ of particle $A$ is in box two.
    \item Anti-haecceitism: There is no difference between these two situations since they are qualitatively identical. This is represented by Z-space, which makes no difference between these two situations: In both cases, there is one particle in box one and in box two.
\end{enumerate}

The reason classical statistical mechanics is relevant for this philosophical debate is, \citet{Huggett1999} claims, precisely the fact that it has two different ways of describing a many-particle system - $\Gamma$-space and Z-space - that correspond to haecceitism and anti-haecceitism, respectively. Therefore, if we are able to find differences in the empirical predictions that $\Gamma$-space and Z-space based descriptions give us, then the metaphysical problem of haecceitism could be settled empirically. In general, there are two questions we need to ask ourselves:

\begin{enumerate}
    \item Is it true that $\Gamma$-space and Z-space entail haecceitism and anti-haecceitism, respectively?
    \item Are there any empirical differences between the predictions of $\Gamma$-space and Z-space?
\end{enumerate}

Regarding the first problem, \citet[p. 410-415]{Gordon2002} has opposed Huggett's view. First, he argues, classical particles can be identified qualitatively via their spatiotemporal location, such that $\Gamma$-space, although compatible with haecceitism, does not require it. Second, the fact that Z-space does not distinguish between configurations in which the positions of qualitatively identical particles are exchanged does not imply that it is anti-haecceistic, this indifference of Z-space can simply be explained by classical ignorance.

Regarding the second problem, Huggett introduces a statistical argument and an argument based on entropy. Here, we discuss the statistical argument, which is developed as follows: We assume that there is a uniform probability distribution over all possible configurations of the system (all possible worlds in the modal universe), such that every configuration has the same probability. In $\Gamma$-space, there are four different setups - both particles on the left, both on the right, $A$ on the left and $B$ on the right and $B$ on the left and $A$ on the right. Every configuration has a probability of $1/4$. In $Z$-space, however, the latter two configurations are identical, such that there are three different options that all have a probability of $1/3$. Consequently, $\Gamma$-space and $Z$-space will make different predictions for the observed frequencies.

\citet{Huggett1999} has argued that this argument is incorrect: If one takes into account that the particles are impenetrable or, less restrictively, that the set of phases with co-located particles has measure zero in phase space, both descriptions will give the same frequencies. Let us assume that each box can only contain up to one particle (we think about the boxes as a discretization of space and assume the particles to be impenetrable). In this case, \ZT{one particle on the left and one on the right} is the only possible configuration, which then has a probability of one in both descriptions. In $\Gamma$-space, this results from the fact that there are two qualitatively identical configurations - one with $A$ on the left and one with $A$ on the right - that are both assigned a probability of $1/2$. A similar argument could be run, for example, if we have two particles and three boxes: $Z$-space then provides us with three configurations that each have a probability of $1/3$, $\Gamma$-space gives six options with probability $1/6$ that come in indistinguishable pairs, giving three distinguishable configurations that also have a probability of $1/3$. Thus, there is, in fact, no difference between $\Gamma$-space and $Z$-space as far as observable frequencies of particle configurations are concerned.

It thus appears as if classical statistical mechanics has no empirical answer to the problem of haecceitism to offer. However, the situation changes if we consider more recent results from density functional theory.

\section{\label{opd}Order-preserving dynamics}
It is important to note that using a field theory does not commit one to assuming that all particles are qualitatively identical. In fact, there is a very straightforward way to avoid this: If our system consists of two different types of particles - let us call them $A$ and $B$ - then we can introduce two dichotomic fields $\rho_A$ and $\rho_B$. The value of $\rho_A(\vec{r})$ is nonzero if and only if there is a particle of type $A$ at position $\vec{r}$, whereas $\rho_B(\vec{r})$ is nonzero if and only if a particle of type $B$ is at position $\vec{r}$. In this way, we are able to distinguish between particles of type $A$ and $B$. Nevertheless, we are still using a field theory. The total density $\rho(\vec{r})$, which tells us whether there is any particle at position $\vec{r}$, is simply given by $\rho = \rho_A + \rho_B$.

What is less straightforward to see, but very important for our line of argument, is that we can also distinguish between different particles if they are all qualitatively identical, that is, if we can only observationally distinguish between them by their positions. In this case, we can pick out a particle, give it the name $A$, and give all other particles the name $B$. We then wish to keep track of the particle $A$. This can be done by again introducing two fields $\rho_A(\vec{r})$ and $\rho_B(\vec{r})$, where $\rho_A(\vec{r})$ is nonzero if and only if the particle with the name $A$ is at position $\vec{r}$ and $\rho_B(\vec{r})$ is nonzero if and only if any other particle is at position $\vec{r}$. Essentially, we are pretending that our system is a mixture of two different particle types, where one component has only one particle and where both particle types have identical properties.

The aforementioned considerations are of profound importance in classical statistical mechanics. To see this, let us consider one of its most important theories, namely dynamical density functional theory (DDFT). DDFT, which was introduced by \citet{Evans1979} based on phenomenological arguments and later derived from the microscopic dynamics of many-particle systems by \citet{MarconiT1999}, \citet{ArcherE2004}, \citet{Yoshimori2005}, and \citet{EspanolL2009}, describes the time evolution of the one-body density $\rho(\vec{r})$, which essentially corresponds to the dichotomic field introduced in \cref{classical}. In the variants of DDFT considered here, $\rho$ is the ensemble average of this dichotomic field, giving the probability of finding a particle at position $\vec{r}$ (from now on, we will always use $\rho$ with this meaning). For an overdamped colloidal $N$-particle system, the dynamics of $\rho$ can be derived by integrating the transport equation for the $N$-body distribution function over the coordinates of all except for one particle. This gives \citep[p. 4248]{ArcherE2004}
\begin{equation}
\frac{\partial}{\partial t}\rho(\vec{r},t)= \Gamma k_B T \Nabla^2 \rho(\vec{r},t) + \Gamma \Nabla\cdot \INT{}{}{r'}\rho^{(2)}(\vec{r},\vec{r}',t)\Nabla U_2(\vec{r}-\vec{r}') + \Gamma \Nabla \cdot \rho(\vec{r},t)\Nabla U_1(\vec{r}),
\label{bbgky}
\end{equation}
where $t$ is the time, $k_B$ the Boltzmann constant, $T$ the temperature, $\Gamma$ the mobility, $\Nabla$ the del operator, $\rho^{(2)}$ the two-body distribution function (giving the probability of finding a particle at $\vec{r}$ and another one at $\vec{r}'$), $U_2$ the interaction potential and $U_1$ the external potential. Equation \eqref{bbgky} is not closed since it involves the unknown function $\rho^{(2)}$. For an equilibrium system, one has
\begin{equation}
k_B T \Nabla \rho(\vec{r}) + \INT{}{}{r'}\rho^{(2)}(\vec{r},\vec{r}')\Nabla U_2(\vec{r}-\vec{r}') =\rho(\vec{r}) \Nabla \frac{\delta F}{\delta\rho(\vec{r})}
\label{equilibrium}
\end{equation}
with the intrinsic free-energy functional $F$. Making the (often reasonably accurate) assumption that \cref{equilibrium} also holds out of equilibrium (\ZT{adiabatic approximation}), one gets (by inserting \cref{equilibrium} into \cref{bbgky}) the DDFT equation \citep[p. 4248]{ArcherE2004}
\begin{equation}
\frac{\partial}{\partial t}\rho(\vec{r},t)= \Gamma \Nabla \cdot \bigg(\rho(\vec{r},t)\Nabla\frac{\delta F}{\delta\rho(\vec{r},t)} + \rho(\vec{r},t)\Nabla U_1(\vec{r})\bigg).
\label{ddftequation}
\end{equation}

Since the early days of DDFT, a variety of extensions (such as inertial dynamics \citep{Archer2006}, particles with general shapes \citep{WittkowskiL2011}, and nonisothermal systems \citep{WittkowskiLB2012}) and applications (such as phase separation \citep{ArcherE2004}, crystal growth \citep{vanTeeffelenLL2008b}, electrochemistry \citep{JiangCJW2014b}, active matter \citep{WittmannB2016}, and disease spreading \citep{teVrugtBW2020,teVrugtBW2020b}) have emerged.

In particular, DDFT is also capable of describing mixtures. This is straightforwardly done in the way described above - for a mixture of two species $A$ and $B$, DDFT has two fields $\rho_A$ and $\rho_B$. Keeping track of the dynamics of a single (\ZT{tagged}) particle has also been a problem of interest in DDFT. \citet{ArcherHS2007} have done this precisely in the way described above: One artificially divides a system of qualitatively identical particles into a mixture where one component consists of only one particle (which has been given a name). This method has been used, e.g., by \citet{BiervRDvdS2008,StopperTDR2018}, and \citet{YeNTZM2019}. An extension of DDFT that is important for our purposes, and that solves certain problems of the standard theory\footnote{DDFT typically employs thermodynamic functionals from grand-canonical statistical mechanics, which is inappropriate if the system has a fixed particle number. PCD avoids this by using canonical methods. This difference between DDFT and PCD can be of practical importance, but is irrelevant for the discussion here.} (but which is identical to DDFT for all purposes relevant to this discussion), is known as particle-conserving dynamics (PCD) and was introduced by \citet{delasHerasBFS2016}. DDFT is a paradigmatic example for many approaches of classical statistical mechanics, which is why \citet{teVrugt2020} has suggested that a more detailed study of DDFT can be beneficial for philosophy of physics. A detailed review of DDFT was given by \citet{teVrugtLW2020}.

In PCD (and in the order-preserving dynamics introduced below), one uses for the free energy $F$ appearing in \cref{ddftequation} the canonical form \citep{delasHerasBFS2016}
\begin{equation}
F = - k_B T \ln Z - \INT{}{}{r}\rho(\vec{r})U_\mathrm{a}(\vec{r})    
\label{freeenergy}
\end{equation}
with the partition function $Z$. Here, $U_\mathrm{a}$ is defined to be the potential for which $\rho(\vec{r})$ is the equilibrium distribution. From \cref{ddftequation,freeenergy}, we find
\begin{equation}
\frac{\partial}{\partial t}\rho(\vec{r},t)=\Gamma \Nabla\cdot  \rho(\vec{r},t)\Nabla(U_1(\vec{r}) - U_\mathrm{a}(\vec{r},t)).
\label{particlecon}
\end{equation}
For mixtures, we get an equation of the form \eqref{particlecon} for each particle species. From \cref{particlecon}, it can be seen that equilibrium is reached if $U_\mathrm{a} = U_1$. Consequently, the way in which $U_\mathrm{a}$ is calculated characterizes the equilibrium state of the theory \citep{WittmannLB2020}. So far, we did not have to make strong restrictions regarding spatial dimension, particle number or interaction potential.

The (thought) experiment we are now concerned with is very simple: We have a system consisting of two hard rods $A$ and $B$ confined to a box in one spatial dimension (1D). What is special about a 1D system is that, if rod $A$ is initially to the left of rod $B$, it will always remain there. For them to switch positions in a 1D system, it would be necessary that rod $A$ passes through rod $B$, which is impossible since they are hard and therefore impenetrable. If we want to describe the dynamics of the particles in this system, we have two options: First, we can work in $\Gamma$-space by solving the equations of motion for the two individual particles that describe their position as a function of time. Second, we can introduce two fields $\rho_A(\vec{r})$ and $\rho_B(\vec{r})$, telling us whether we will find rod $A$ or rod $B$, respectively, at position $\vec{r}$.

Both options have been used in an interesting article by \citet{SchindlerWB2019}, where the system of rods in 1D described above was explored. First, they performed Brownian dynamics simulations\footnote{In a Brownian dynamics simulations, one solves the equations of motion for all individual particles. Technical details on the specific simulations described here can be found in the appendix of \citet{SchindlerWB2019}.}, which (as is common in soft matter theory) are treated as an exact reference. This gives the correct result: If rod $A$ is initially to the left of rod $B$, solving the equations of motion of $\Gamma$-space for them never gives a situation where rod $A$ is on the right of rod $B$. For the field-theoretical description, \citet{SchindlerWB2019} derived an extension of PCD for mixtures that allows for a very accurate description (the adiabatic approximation that is made compared to the $\Gamma$-space simulations has only small effects). However, when using the second type of description - two fields $\rho_A$ and $\rho_B$ measuring the positions of rods $A$ and $B$ - a strange observation was made: Starting from a configuration with rod $A$ on the left of rod $B$ (as indicated by the peaks of the corresponding density fields), this theory also leads to solutions where rod $A$ is on the right of rod $B$ (as apparent from the nonzero value of $\rho_A$ at positions where the probability to find particle $A$ should be zero). Consequently, PCD for mixtures does not respect the conservation of particle order.

This effect is a consequence of the underlying statistical description: DDFT and related theories assume that the system is ergodic, that is, that it explores all possible states. In the description of hard rods, one uses an interaction potential that is infinite if two particles overlap and zero otherwise. Consequently, the theory does not allow for configurations where particles overlap (energetically impossible), but it does allow for any other configuration, including those where particles switch their positions. 

\citet{WittmannLB2020} have derived an improved theory, which they refer to as \ZT{order-preserving dynamics} (OPD). Here, the statistical theory describing two particles $A$ and $B$ (where $A$ is initially on the left) is based on an asymmetric interaction potential that is zero if particle $A$ is on the left of particle $B$ and infinite otherwise. In this way, the particle order is taken into account in the statistical average. Note that the actual physical interaction potential does, of course, not depend on the particle order. Instead, the asymmetric interaction potential has the effect that otherwise identical configurations with particle $A$ on the left and with particle $B$ on the left get assigned different probabilities (namely, the latter case gets probability zero). Numerical calculations based on OPD rather than PCD show that OPD respects the conservation of particle order (as it is supposed to), that is, the field $\rho_A$ always vanishes at positions that particle $A$ cannot reach.

Mathematically, PCD assumes (for a system of two hard rods of length $2R$ in one dimension) that in equilibrium the probability for a configuration with particle $A$ at $x_1$ and particle $B$ at $x_2$ is\footnote{This holds for $x_1,x_2 \in [R,L-R]$ if $x_1$ and $x_2$ are the center-of-mass positions of the rods (assuming them to have a homogeneous mass density). Other values of $x_1$ and $x_2$ are always excluded by the infinite potential describing the system's walls.} given by $p = e^{-\beta U_2^H(|x_2 - x_1|)}/(2!\Lambda^2 Z_2)$, where the partition function $Z_2 = (2!\Lambda^{2})^{-1}\TINT{R}{L-R}{x_1}\TINT{R}{L-R}{x_2}e^{-\beta U^H_2(|x_2 - x_1|)}$ is a normalization, $\Lambda$ is the thermal de Broglie wavelength, $\beta = (k_B T)^{-1}$ is the thermodynamic beta and 
\begin{equation}
U^H_2(|x|) = 
\begin{cases}
0 & \text{ for }|x| > 2R,\\
\infty & \text{ otherwise }
\end{cases}
\label{hardcore}
\end{equation}
is the hard-core interaction potential. This gives \citep{SchindlerWB2019}
\begin{equation}
\begin{split}
&\rho_A(x)=\rho_B(x)=\INT{R}{L- R}{x_1}\INT{R}{L- R}{x_2}\frac{e^{-\beta U^H_2(|x_2 - x_1|)}\delta(x-x_1)}{2! \Lambda^2 Z_2}=\INT{R}{L- R}{x_1}\INT{R}{L- R}{x_2}\frac{e^{-\beta U^H_2(|x_2 - x_1|)}\delta(x-x_2)}{2! \Lambda^2 Z_2}\\ =
&\frac{1}{(L-4R)^2}((L -3R - x)\Theta(x - R)\Theta(L - 3R - x) + (x -3R)\Theta(x - 3R)\Theta(L - R - x)),   
\end{split}
\label{symmetric}
\end{equation}
where $L$ is the length of the system and $\Theta$ the Heaviside stepfunction. Notably, the equilibrium profiles $\rho_A$ and $\rho_B$ are identical, which is unphysical. On the other hand, if we use the ordering potential
\begin{equation}
U^O_{2}(x) = 
\begin{cases}
0 & \text{ for }x > 2R,\\
\infty & \text{ otherwise },
\end{cases}
\label{orderingpotential}
\end{equation}
we get \citep{SchindlerWB2019}
\begin{align}
&\rho_A(x) =\INT{R}{L-R}{x_1}\INT{R}{L-R}{x_2}\frac{e^{-\beta U^O_2(x_2 - x_1)}\delta(x-x_1)}{\Lambda^2 Z_{11}} = \frac{2(L -3R - x)\Theta(x - R)\Theta(L - 3R - x)}{(L-4R)^2},\label{asymmetric1}\\
&\rho_B(x)=\INT{R}{L-R}{x_1}\INT{R}{L-R}{x_2}\frac{e^{-\beta U^O_2(x_2 - x_1)}\delta(x-x_2)}{\Lambda^2 Z_{11}} = \frac{2(x -3R)\Theta(x - 3R)\Theta(L - R - x)}{(L-4R)^2}\label{asymmetric2}
\end{align}
where the ordered partition function $Z_{11} = \Lambda^{-2}\TINT{R}{L-R}{x_1}\TINT{R}{L-R}{x_2}e^{-\beta U^O_2(x_2 - x_1)}$ is again a normalization. Unlike in \cref{symmetric}, the particle order is now respected. For out-of-equilibrium systems, both PCD and OPD predict an approach to equilibrium. Since the equilibrium state is ordered in OPD, but not in PCD, this explains why PCD predicts an unphysical mixing.

Let us now introduce OPD in a more general form following \citet{WittmannLB2020}. We consider a system of $N$ particles in one dimension and divide it into three \ZT{species}: one \ZT{tagged} particle (T) at position $x_0$, the $N_L$ particles on the left of the tagged particle, and the $N_R$ particles on its right. The density profiles then read
\begin{equation}
\rho^{\nu}(x)=\INT{}{}{x_0}\frac{Z_{N_L}^L(x_0)Z_{N_R}^R(x_0)}{\Lambda Z_N}e^{-\beta U_1^T(x_0)}\rho^{\nu}(x|x_0)
\label{conditional}
\end{equation}
with $\nu \in \{L,T,R\}$. Here, $U_1^T$ is the external potential acting on the tagged particle,
\begin{equation*}
Z_{N_\alpha}^\alpha(x_0)=\INT{}{}{x^{N_\alpha}}\frac{e^{-\beta (U_{N_\alpha} + V_{N_\alpha}+W_{N_\alpha})}}{N_\alpha!\Lambda^{N_\alpha}}
\end{equation*}
with $\alpha \in \{L,R\}$ is the conditional partition function, $Z_N$ is the ordinary partition function,
\begin{equation}
\rho^{\nu}(x|x_0) = \INT{}{}{}{x^{N_\alpha}}\frac{e^{-\beta (U_{N_\alpha} + V_{N_\alpha} + W_{N_\alpha})}}{N_\alpha! \Lambda^{N_\alpha}Z^\alpha_{N_\alpha}(x_0)}\hat{\rho}_\alpha(x) 
\label{zwei}
\end{equation}
(with $\rho^{T}(x|x_0) = \delta(x-x_0)$) is the conditional density, $\hat{\rho}_\alpha(x) =\sum_{i=1}^{N_\alpha}\delta(x - x_i)$ with the position $x_i$ of the $i$th particle is the density operator for species $\alpha$, $U_{N_\alpha}$ is the interaction potential for the particles of species $\alpha$ when ignoring the tagged particle, $V_{N_\alpha}$ the external potential for species $\alpha$, and
\begin{equation}
W_{N_\alpha} = \sum_{i=1}^{N_\alpha}w_\alpha(x_i - x_0) 
\end{equation}
is the potential for interactions with the tagged particle, constructed from the order-preserving potentials
\begin{equation}
w_\alpha(x_i - x_0) =
\begin{cases}
0 & s_\alpha(x_i - x_0 ) > 2R,\\
\infty & s_\alpha (x_i - x_0)< 2R,
\end{cases}
\label{casedist}
\end{equation}
where $s_L = -1$ and $s_R = +1$. Intuitively, \cref{conditional} can be understood in a Bayesian way - the probability $\rho^\nu(x)$ of finding a particle of species $\nu$ at position $x$ is given by the probability of finding it at $x$ given that the tagged particle is at position $x_0$ (this is what is measured by the conditional density $\rho^{\nu}(x|x_0)$) multiplied by the probability that the tagged particle is at $x_0$ (proportional to $e^{-\beta U_1^T(x_0)}$). Moreover, the conditional probability $\rho^{\nu}(x|x_0)$ is, according to \cref{zwei}, given by an average over all possible configurations of the system (determined by all particle positions). The probability of each configuration is finally proportional to $e^{-\beta (U_{N_\alpha} + V_{N_\alpha}   + W_{N_\alpha})}$. Due to \cref{casedist}, this probability is zero for \ZT{forbidden} particle configurations. Equation \eqref{conditional} can then be used to calculate, from the density profiles $\rho^\nu(x)$, the potential $U_a$ required in \cref{particlecon} by the algorithm \citep{delasHerasBFS2016}
\begin{equation*}
\beta U_{a,n}^\nu (x) = \beta U_{a,n-1}^\nu(x) - \ln\rho^\nu(x) + \ln \rho^\nu_{n-1}(x),    
\end{equation*}
where in each step $\rho^\nu_{n-1}$ is calculated from the corresponding potential $U_{a,n-1}^\nu$ via \cref{conditional}. From now on, we restrict ourselves to the two-particle case. Here, it is irrelevant whether one tags the left or the right particle. The first case would correspond to $\rho_A = \rho^T$ and $\rho_B = \rho^R$ with $\rho^L=0$, the second one to $\rho_A = \rho^L$ and $\rho_B = \rho^T$ with $\rho^R=0$.

A general physical motivation for the development of OPD (apart from the intention to describe the rather simple setup above) is the description of nonergodic systems, in particular of glasses. Roughly speaking, glass formation in a dense liquid occurs if the liquid is cooled below its freezing temperature, but cannot freeze because the particles get trapped and cannot move to their equilibrium positions in a crystal. Therefore, \citet{SchindlerWB2019} argued, DDFT is incapable of describing glass formation if it allows hard particles to pass through each other, since this makes caging effects relevant for glass formation impossible. OPD, which gets rid of this problem, is intended to be a first step towards a description of caging in DDFT. See \citet{teVrugtLW2020} for a more sophisticated discussion of the relation between DDFT and the glass transition.

\section{\label{haecc2}The case for haecceitism}
Having introduced the physics of OPD, we can now turn to its philosophical implications: Which new insights does it provide on the problem of haecceitism compared to older statistical theories? As discussed in \cref{classical}, we have to distinguish between two different questions: First, it has to be clarified whether different physical descriptions have different metaphysical implications regarding the problem of haecceitism, and second, we need to analyze whether there are empirical differences between them. If both questions are answered with \ZT{yes}, it is possible to give an empirical answer to the metaphysical question whether haecceitism is correct. Typically, the two physical descriptions that are compared are $\Gamma$-space and Z-space. However, as the discussion in \cref{opd} should have made clear, we can also compare different types of field theories, in particular PCD and OPD. Thus, we re-formulate the problems from \cref{classical} as

\begin{enumerate}
    \item Do OPD and PCD entail haecceitism and anti-haecceitism, respectively?
    
    \item Are there any empirical differences between OPD and PCD?
\end{enumerate}

Whether PCD is anti-haecceistic is not really clear. PCD allows to follow the trajectory of an individual particle. As far as this point is concerned, it is similar to a $\Gamma$-space description in that the individuation is done (or at least can be done) via the spatiotemporal trajectory, which is a qualitative criterion for individuation that is therefore compatible with anti-haecceitism. Consequently, there are good reasons to believe that the relation of PCD to haecceitism is similar to $\Gamma$-space's relation to haecceitims, namely neutral. Nevertheless, if it turns out that OPD is more accurate than PCD, and if this is a consequence of haecceistic properties of OPD, we still have a good argument for haecceitism.

And there are good reasons to believe that OPD is a haecceistic theory. It treats situations with particle $A$ on the left as being different from situations with particle $B$ on the left\footnote{As discussed in \cref{opd}, OPD assigns different probabilities to these configurations, and this only makes sense if these (qualitatively identical) configurations are different.}. Since $A$ and $B$ are qualitatively identical, the only difference between these situations is what they represent de re about particle $A$, that is, we have a haecceistic difference by the definition introduced in \cref{haecc1}. However, one might object that the fact that OPD is designed to ensure that particle $A$ is always on the left of particle $B$ does not imply that it is haecceistic since particle $A$ is just defined as \ZT{the particle that is initially on the left}. This is a qualitative criterion that allows one to distinguish particle $A$ from particle $B$, and treating $A$ and $B$ differently based on a qualitative difference (their initial position) is perfectly compatible with anti-haecceitism.

To understand this issue in more detail, we should remember the discussion of Kripke's designation theory from \cref{haecc1}, where we have introduced the distinction between names and definite descriptions. In OPD and PCD, particles are given indices, such that a certain particle is referred to as, for example, \ZT{particle $A$}. These indices can be understood in two ways: First, we can interpret \ZT{particle $A$} as standing for \ZT{the particle that is initially on the left}. In this case, \ZT{particle $A$} is a definite description (like \ZT{the winner of the election}), since particle $A$ is identified via its qualitative properties. Second, we can interpret \ZT{particle $A$} as a name (such as \ZT{Nixon}), that serves as a rigid designator across possible worlds. This is also possible for both haecceists and anti-haecceists, since both can admit that the particles have qualitative thisness (which, as discussed at the end of \cref{haecc1}, allows for naming). However, only haecceists can then stipulate two different possible worlds that differ only in the fact that particles $A$ and $B$ are exchanged. As we will show now, this is precisely what is done in OPD.

The reason is that OPD is a statistical theory that describes, like many theories in statistical mechanics, ensemble averages. An ensemble is a collection of infinitely many copies of a system that are in different states. Associated with this ensemble is a probability distribution giving the probability that a system chosen at random from the ensemble is in a certain state. It is a basic principle of statistical mechanics\footnote{See \citet{FriggW2020} and \citet{Reichert2020} for a critical discussion of this principle.} that the time average of an observable (which is what we actually observe in experiments) is, under certain conditions, the same as ensemble average of this observable. This holds if the system is ergodic \citep{FriggW2020}. Since the one-dimensional system violates ergodicity, OPD modifies the standard canonical distribution by replacing the physical interaction potential \eqref{hardcore} by the ordering potential \eqref{orderingpotential}. Thereby, the ensemble averages in \cref{asymmetric1} and \cref{asymmetric2} correctly reproduce the observed density profiles.

From a philosophical perspective, ensemble averages can be thought of as averages over (a set of) possible worlds - actually, \citet[pp. 15-20]{Kripke1980} and \citet[p. 9]{Huggett1999} characterize \ZT{possible worlds} simply as possible states of the world. The possible worlds considered in an ensemble average arise from Kripkean stipulation. As can be seen from \cref{symmetric,asymmetric1,asymmetric2}, the ensemble considered in PCD and OPD contains worlds with $A$ on the left and worlds with $A$ on the right. OPD in particular assigns a different probability to these two different types of worlds, implying that they must be different. Suppose now that \ZT{particle $A$} is a definite description. In this case, there would have to be a qualitative criterion that allows us to pick out a certain particle in one of the possible worlds as particle $A$. Since particles $A$ and $B$ have all their intrinsic properties in common, the only way to pick out particle $A$ is the position, in particular the fact that $A$ is the particle that, in our lab or computer experiment, is on the left. If we interpret \ZT{particle $A$} as \ZT{the particle that is on the left}, then there cannot be a possible world in which particle $A$ is on the right. However, the ensemble does contain such worlds. Hence, assuming that \ZT{particle $A$} is a definite description leads to a contradiction, implying that it must be a name.

Kripkean stipulation is only possible for anti-haecceists if the de re specification can at least in principle be replaced by a qualitative one \citep[p. 222]{Lewis1986}. Here, this replacement is not possible since the situations with particle $A$ and with particle $B$ on the left are qualitatively indistinguishable. Hence, in OPD, we require Kripkean stipulation of a type that is only available to the haecceist. In particular, we require that it is possible to identify a certain particle as \ZT{particle $A$} in different worlds from the modal universe, that is, that we can point at the particle that is on the left in our world and say that this specific particle could have been on the right. Hence, if it turns out that we require OPD rather than PCD for an accurate description of the outcome of experiments, then we have an empirical argument for haecceitism and for the existence of modality de re!

At this point, the careful reader might raise the following point: We have claimed that we cannot individuate particle $A$ as \ZT{the particle that started on the left} since the ensemble average considers all possible states, including those where it is on the right. However, in OPD configurations with $A$ on the right are excluded, so there is in fact only one possibility for the particle order. Two things can be said in response. First, when looking at the way average values are computed in \cref{symmetric} (PCD) and \cref{asymmetric1,asymmetric2} (OPD), one can see that the calculation is in fact completely analogous, that is, one integrates over the full phase space (all possible configurations) in both cases. The only difference is the probability assigned to different states, determined by the interaction potential that is employed. This probability is - in OPD - zero for excluded configurations. One could now (as pointed out in \citet{SchindlerWB2019,WittmannLB2020}) simply reduce the phase space and integrate only over the \ZT{allowed} configurations. In this (mathematically equivalent) framework, only different configurations with the same particle order (that all have non-haecceistic differences) are considered. However (and this is the second point), this approach only works as long as the probability of exchanged configurations is exactly zero. DDFT applies to more general setups than just one-dimensional two-particle systems, and a crucial motivation for the development of OPD-like theories is the description of glassy systems in three dimensions. In such a system, an exchange of particles $A$ and $B$ is possible, but it takes a very long time due to caging. This suggests that, in a (yet to be developed) extension of OPD to higher dimensions, configurations with $A$ and $B$ exchanged should be assigned a small but nonzero probability at short and intermediate times. In this case, it is no longer possible to consider just one possibility, one has to consider all (haecceistically differing) configurations.

One can also develop a second argument for a haecceistic interpretation of the configuration we consider. The two-particle-system considered here has a certain similiarity to a thought experiment by \citet{Black1952}, originally developed to discuss the identity of indiscernibles (a principle that, as mentioned above, is also relevant here as it is violated in haecceitism). Imagine a universe that contains nothing but two spheres at a certain distance that have all qualitative properties (size, mass, color, ...) in common. If we assume that the identity of indiscernibles holds, then - given that there is absolutely no way to distinguish the spheres - it is difficult to defend the position that we have two spheres. One might now object that they can be distinguished based on their spatial position. However, even this position is, strictly speaking, only reasonable if we allow \ZT{positions} to have an independent existence. Otherwise, the only thing one can say is that the spheres have a distance from each other (which does not allow to distinguish them in any way), and the only thing that allows one to distinguish the places where the spheres are is that different spheres are in this place (which presupposes that they are different) \citep[p. 158]{Black1952}. 

Thus, even the assumption made above that we can distinguish particles based on their position in the absence of other distinguishing features (an assumption that it quite common in the literature on this topic) might have been too optimistic. Although the situation is a little easier for the setup considered here since our one-dimensional system is finite (that is, it has walls which allow one to distinguish positions based on their distance from a wall), it still arises in quite a similar way if we assume that both particles are at the same distance from a wall. Note that nothing allows for a distinction between the different(?) walls! If we are haecceitists, however, things are a lot easier. In particular, if we defend a form of haecceitism that includes the existence of haecceities attributable to the two particles (see \cref{haecc1}), these haecceities provide a solid basis for giving the particles names - such as $A$ and $B$. We can then introduce a coordinate system, for example by calling the position of the wall closer to $A$ (this picks out a unique position) $\vec{r} = \vec{0}$. From then on, it makes perfect sense to stipulate a situation in which the two particles are exchanged.

An important difference between Black's thought experiment and OPD is that the former assumes the two spheres to be the only objects in the universe, whereas OPD (in its current form) is set up in the (grand-)canonical ensemble where particles are in contact with a heat bath. The reason is that colloidal particles in a fluid are DDFT's main field of application. However, the crucial point of Black's argument is that the two-sphere-universe is entirely symmetric, such that the spheres have no properties that allow to distinguish them. If we assume (as is common) that the fluid acting as a heat bath for the colloids is uniform, a universe consisting only of two colloids at rest in this fluid would still provide no qualitative basis for distinguishing between them. Moreover, it is also conceivable to formulate OPD in the microcanonical ensemble, in which case there would be no heat bath.

Regarding the second question (whether PCD and OPD make different predictions), the fact that there are empirical differences is precisely the reason OPD was developed: PCD predicts that particles can switch their positions, whereas OPD conserves particle order. Consequently, there is an empirical difference between PCD and OPD, and OPD is correct since hard rods in one dimension cannot pass through each other. Although the world we live in is, of course, not one-dimensional, experiments on one-dimensional particle motion (\ZT{single-file diffusion}) can be and have been performed \citep{GuptaNMD1995,HahnKK1996,LinMCRD2005}. For example, \citet{HahnKK1996} have studied molecules with a diameter of 0.47 nm confined to a channel of diameter 0.73 nm, creating a setup in which position-switching is not possible.

One could now argue that the argument is circular: The fact that PCD does not predict the conservation of particle order is only problematic if we assume that particle order should be conserved. To put it differently: If we assume that it does not make any physical difference if particles $A$ and $B$ are exchanged, then the fact that PCD allows for an exchange of particles $A$ and $B$ does not imply that it makes unphysical predictions. The argument for haecceitism from OPD would then only be convincing if we are already convinced that exchanging $A$ and $B$ makes a difference. However, this argument does not go through since, as discussed above, the spatiotemporal trajectory of an impenetrable particle is a qualitative way of individuating it, available also to the anti-haecceist. Consequently, the predictions of PCD for the trajectory of an individual particle can be tested\footnote{The problems with identification via spatiotemporal position discussed by Black are not relevant here, since an experimental test requires the presence of an observer, who can label one of the particles and thereby distinguish them \citep[p. 157]{Black1952}.} (and found to be false) objectively, without requiring any prior assumptions about haecceitism.

Moreover, the empirical differences between PCD and OPD are not restricted to problems with particle order: To see this, note that in both theories we can consider the sum $\rho = \rho_A + \rho_B$ of the fields $\rho_A$ and $\rho_B$, which gives the probability of finding any particle at position $\vec{r}$. Regardless of one's position on (anti-)haecceitism, it certainly does make an empirical difference if PCD and OPD make different predictions for the field $\rho$ in single-file diffusion experiments or glass-forming systems.

Very remarkably, PCD and OPD do not only disagree on their predictions for the fields $\rho_A$ and $\rho_B$, they also disagree on their predictions for the total density field $\rho$. A good example is figure 6 in \citet{WittmannLB2020}, which compares density profiles for two hard rods in a one-dimensional system. In PCD, the field $\rho$ has larger values at the boundary of the system, meaning that it predicts a larger probability to find any particle close to the boundary than OPD. These results can be compared to microscopic simulations, which model the dynamics of the individual particles by solving the Langevin equations governing their motion. Since these Langevin equations are extremely well understood and confirmed, it is common in soft matter physics to use them as a reference to test field theories \citep{teVrugtLW2020,BickmannBJW2020}. The simulations confirm the predictions of OPD for the density at the boundaries. Moreover, experimental results are available for the evolution of the mean-squared displacement $\braket{x^2}$ in single-file diffusion, which is proportional to $t^{1/2}$ \citep{HahnKK1996}. Comparing these to the predictions of PCD ($\braket{x^2} \propto t$) and OPD ($\braket{x^2} \propto t^{2/3}$) also shows that ODP is more accurate than PCD. The remaining differences between OPD and experiment can be attributed to the employed approximations \citep{WittmannLB2020}.

Our analysis of the role of ensemble averages in OPD has thus shown that, in contrast to what is claimed in the literature, classical statistical mechanics does allow to make a strong case for haecceitism. While we can individuate classical particles with the same intrinsic properties within our world qualitatively based on their spatial location, we can also individuate qualitatively indistinguishable particles across the modal universe based on their primitive thisness. This allows to construct statistical ensembles via Kripkean stipulation in which qualitatively identical configurations are given different statistical weights, something which - although this may seem counterintuitive - is required for a correct description of one-dimensional diffusion.

Three possible objections should be considered at this point. First, it can be argued that there might be a different (yet to be developed) theory from statistical mechanics which is empirically equivalent to OPD, but anti-haecceistic. At present, it is not clear whether this is the case, and it is difficult to see how this theory should be set up given the central role of particle labels in OPD. Nevertheless, such a hypothetical theory would of course force us to reconsider some of the arguments developed here. Let us consider in particular the adiabatic approximation introduced in \cref{opd}. \citet{WittmannLB2020} have attributed the differences in the total density profile $\rho$ between PCD and OPD to the different ways in which the adiabatic approximation acts on the equilibrium framework of these theories. It might thus be possible that these issues are resolved in a theory incorporating also superadiabatic effects (such as the power functional theory (PFT) developed by \citet{SchmidtB2013}), although this is a calculation that is yet to be done. However, PCD and OPD also give different results for the equilibrium profiles of $\rho_A$ and $\rho_B$ (as shown in \cref{symmetric,asymmetric1,asymmetric2}), and the adiabatic approximation is exact (that is, not an approximation) for the equilibrium case. Consequently, even if a PFT based on a symmetric interaction potential gives the correct stationary distributions, it would interpret this distribution not as \ZT{equilibrium}, but as \ZT{a nonequilibrium steady state in which the system remains permanently due to superadiabatic forces}. Since it is a general principle of thermodynamics that isolated systems move to equilibrium and remain there \citep{BrownU2001}, it is reasonable to interpret the stationary state that our one-dimensional system spontaneously enters as an equilibrium state, which is only possible in OPD.

Second, one could object that our world is not one-dimensional, such that results resting on the peculiarities of a one-dimensional system are irrelevant for the real world. However, as discussed in \cref{opd} and in \citet{SchindlerWB2019}, OPD is motivated by the occurence of closely related problems in real three-dimensional systems (namely in glasses). The problems that arise when trying to model the glass transition in DDFT - namely the failure of caging due to particles passing through each other - are the same that arise in the one-dimensional diffusion experiment. Consequently, an extension of OPD to three dimensions (which is yet to be derived) is likely to provide an empirically adequate model of glasses.  Moreover, as discussed above, there are experimental studies of one-dimensional diffusion. For these experiments, OPD provides a better description than PCD. Even though, of course, single-file diffusion is a rather specific experiment, it already provides a strong argument for haecceitism if there is one experiment that is not appropriately explained with an anti-haecceistic theory. Unless we assume that haecceitism only holds for particles in thin channels (which is a rather implausible metaphysical position), the fact that haecceitism holds for particles in thin channels allows to infer that it holds in general.

Third, there are more general philosophical objections to haecceitism (which are not related to this particular example). Since Lewis' definition of haecceitism has been our starting point, we here address his anti-haecceitism. It is based on his counterpart theory introduced in \cref{haecc1}. As discussed by \citet[p. 102]{Weatherson2015}, it is difficult to make sense of haecceistic differences in counterpart theory: Consider a world exactly like ours, with the only difference (a haecceistic one) being that Barack Obama and Julius Caesar have changed roles (e.g., Obama has counquered Gaul). This would require that the Caesar of this world is the counterpart of the Obama of our world, an assumption for which there is absolutely no reason given that this world is qualitatively identical to ours. \citet[p. 231]{Lewis1986} considers as an example the possibility that he was born as one of two twins. On a haecceistic understanding, there would be two possible worlds, one in which he was the first-born and one in which he was the second-born twin. Lewis, in contrast, argues that these are two possibilities within a single world since this world contains two counterparts for him (the two twins).

Let us now try to make sense of OPD within counterpart theory. Our ensemble average involves pairs of configurations differing only in which particle is on the left. For Lewis, these cannot correspond to different possible worlds since they are qualitatively identical. Instead, we have to think about \ZT{$A$ on the left} and \ZT{$B$ on the left} as two different possibilities within a single world. This is analogous to the twin example, where \ZT{first-born twin} and \ZT{second-born twin} are two different possibilities within a single world. Regarding the possibility that he might have been someone else, \citet[p. 232]{Lewis1986} emphasizes that it is \ZT{a possibility for me, not for the world}. Similarly, the particle on the left and the one on the right are possible ways for particle $A$ to be.

The problem with this view in the context of statistical mechanics is that \ZT{possible ways for the world to be} is what elements of a statistical ensemble are taken to represent. Probabilities in an ensemble average are thus assigned not to possibilities for a certain particle, but to possibilities for the world. Let us assume, for clarification, that the world only consists of these two particles. Each possibility for the world to be is a possible world. We are considering now the possibility \ZT{$A$ is at position $x_1$ and $B$ is at $x_2$}. This is a full specification of the possible world. Next, we consider the possibility \ZT{$B$ is at $x_1$ and $A$ is at $x_2$} and assign it (as required by OPD) a different probability, meaning that it must be a possibility different from the first one (otherwise it could not have a different probability). But if the possibilities are different and if each possibility is a possible world, then the possible worlds must be different. However, they are qualitatively identical. Lewis' counterpart theory, while it can accommodate thinking of the particle on the right in a possible world as a possible way for particle $A$ to be, it cannot accomodate thinking of two qualitatively identical worlds as being different. This is something that (by definition) only haecceitism is capable of.

\section{\label{gibbs}The inverse Gibbs paradox}
Finally, we relate our results to the famous Gibbs paradox (see  \citet{Saunders2013,Saunders2018}). Consider two boxes filled with gas that are separated by a wall. When the wall is removed, the gases mix, which gives rise to an entropy increase (\ZT{mixing entropy}). This mixing entropy is present even for very similar particles, but vanishes as soon as the particles are completely qualitatively identical. In classical statistical mechanics, this behavior can be recovered by thinking of two phases differing only by the exchange of indistinguishable particles as identical. Frequently, the Gibbs paradox is assumed to be a foreshadowing of quantum mechanics.

The mixing paradox was also discussed in the context of PCD for two particles in one dimension by \citet{SchindlerWB2019}. They constructed a so-called \ZT{inverse mixing paradox}: Consider two hard rods $A$ and $B$ in one dimension that are initially in physical equilibrium. The distributions of $\rho_A$ and $\rho_B$ in physical equilibrium have to respect particle order. If one then studies the time evolution of the system using PCD, it is found that the system moves out of the physical equilibrium state and evolves into an unphysical equilibrium state in which the distributions $\rho_A$ and $\rho_B$ are identical. The reason is that, if one describes particles $A$ and $B$ (which have the same intrinsic properties) as two different species, the total entropy increases by the mixing of particles $A$ and $B$ due to the mixing entropy, and since PCD allows the particles to exchange positions, it therefore predicts that the system evolves towards this maximal-entropy state. \citet{SchindlerWB2019} then argue that the analogy to the mixing paradox is that the impenetrability of the particles that prevents them from switching positions in reality plays the role of a hard wall. The actual maximum of the entropy is the demixed state, whereas the entropy of the mixed state is not defined.

Regarding the standard Gibbs paradox, this result speaks in favour of explaining the absence of a mixing entropy for qualitatively indistinguishable particles simply by the fact that the definition of entropy depends on the experimenter's choice of thermodynamic variables \citep{Jaynes1992}. If we would consider the same setup in two dimensions, then it would typically be correct that the particles switch positions after some time, and the reason we do not usually attach an entropy increase to this process is only that it is of no relevance to thermodynamic experiments.

Nevertheless, the inverse mixing paradox shows that there is, of course, an important difference between the two-dimensional case (where mixing is possible) and the one-dimensional case (where it is not). In one dimension, the difference between the particles matters, whereas in two dimensions it is very natural to ignore these differences such that there is no mixing entropy. This can be understood using the concept of demarcating properties used by \citet{Saunders2018} in his analysis of the Gibbs paradox. The idea is that, in statistical mechanics, distinguishability is a non-fundamental property. A set of indistinguishable particles may have properties arising from their dynamics that allow for an effective description within a reduced phase space in which they are distinguishable. As an example, \citet{Saunders2018} considers a set of coins with identical intrinsic properties in a box, some of which show heads and some tails. If the box is not violently shaken, the coins will not flip. Consequently, despite the fact that (a priori) all coins are alike, the property \ZT{showing heads} or \ZT{showing tails} is dynamically stable, and a reduced description can be developed in which the coins are divided into two species based on which side they show.

This example has a striking similarity to the case of hard rods in one dimension discussed here. Just like the (initially qualitatively indistinguishable) coins become distinguishable by showing either heads or tails and being unable to change this, the (initially qualitatively indistinguishable) rods become distinguishable by being either on the left or on the right and not being able to pass each other. \citet{Saunders2018} even mentions that in \ZT{special cases} (but only then), the place of origin can be a demarcating property. Here, we have such a case.

Note, however, that the fact that the demarcating properties allowing for observational distinguishability arise only as a limiting case does not affect the validity of our arguments in favour of haecceitism from the previous section. If, in Saunder's example, we have a coin $A$ and a coin $B$ showing heads and tails, respectively, we can construct by Kripkean stipulation a world that is identical to ours apart from the fact that $A$ shows tails and $B$ shows heads. In the construction of an OPD-like theory for the coin system, this configuration, which is qualitatively identical to the actual one, would then have to be assigned a different probability, which is only possible by appealing to the haecceity of the coins. The important point here is that, to use the terminology introduced in \cref{haecc1}, conceptual distinguishability may be present at the fundamental level (through haecceities), whereas observational distinguishability is non-fundamental. 

\section{\label{qm}But what about quantum mechanics?}
In this article, we have established an argument in favour of haecceitism on the basis of a scientific theory, in this case OPD (which belongs to classical statistical mechanics). This approach belongs to the tradition of inductive metaphysics \citep{Scholz2019}, which attempts to answer metaphysical questions in an empirically (in particular: scientifically) informed way. Inductive metaphysics has been employed in a variety of contexts, ranging from quantum mechanics \citep{Naeger2020} over thermodynamics \citep{teVrugt2021} to the social sciences \citep{Scholz2018}. This shows already that it is not (and presumably should not be) restricted to \ZT{fundamental} sciences. \ZT{Fundamentality} is a criterion that quantum mechanics presumably satisfies, thermodynamics maybe, and social sciences certainly not. This is important here since it might be questioned whether classical statistical mechanics - a non-fundamental science - is a reasonable basis for metaphysical arguments. The difficulty is that  \ZT{classical statistical mechanics implies haecceitism} is, of course, a different statement than \ZT{haecceitism is true}, and the analysis of OPD presented in this work provides an argument for the former statement. To infer from this the truth of haecceitism requires, we have to assume that (a) classical statistical mechanics provides an accurate description of the world and (b) that metaphysical assumptions involved in successful scientific theories are true. Whether assumption (b) holds is a matter of philosophical debate - although assuming it is likely to be a necessary ingredient of scientific approaches to metaphysics - whereas assumption (a) only holds if quantum effects can be ignored. Even granting (b), it might therefore be the case that haecceitism is not true if quantum mechanics provides evidence for a different position. There are good reasons to assume that quantum mechanics does not support haecceitism (see \citet{DallaChiaraTDF1993} and \citet{RedheadT1991} for a discussion of labels/names in quantum mechanics, and \citet{Saunders2013} and \citet{Maidens1998} for a discussion of haecceitism in this context).

Two things could then be said in response. The first one is that it is already an important insight if classical statistical mechanics implies haecceitism, since this also provides insights into what is (and is not) novel about the metaphysical implications of quantum mechanics. This, in fact, has been an important motivation for previous work on this subject \citep[p. 23]{Huggett1999}. It is often considered an important novel insight from quantum mechanics that particles are, in some sense, not individuals. Therefore, if it turns out that this is already the case in classical mechanics, then quantum mechanics would not be innovative in this regard. In this work, however, we have shown that individuality is very essential for classical statistical mechanics, from which it follows that non-individuality would be an important innovation of quantum mechanics. Black's system of two indistinguishable classical spheres is then very different from a two-electron-system, even though in both cases there will be no observable consequences of a particle exchange.

The second and certainly more controversial response would be that metaphysical consequences of a non-fundamental scientific theory can still be true if we restrict ourselves to its domain of application \citep{Needham2013,teVrugt2021}. Thus, if OPD is (to a very good approximation) true for classical colloidal particles, then it might also follow that haecceitism is true for these particles, even though it is not true for their microscopic quantum-mechanical constituents. In this case, we might even strengthen our statement from \cref{gibbs}, namely \ZT{observational distinguishability is a non-fundamental property}, to \ZT{conceptual distinguishability is a non-fundamental property}. Here, this property emerges during the classical-quantum transition. There is no room here to defend this (certainly interesting and controversial) view, so your takeaway from this article should be the following: OPD shows that classical statistical mechanics requires haecceitism, which at least shows that quantum statistical mechanics is innovative if it does not require haecceitism. It might even show that haecceitism holds for (approximately) classical particles such as colloids, though this depends on whether one is willing to accept a \ZT{non-fundamental haecceitism}.

\section{\label{conclusion}Conclusion}
In order-preserving dynamics, qualitatively identical configurations that only differ in what they represent de re about a certain particle are treated differently in the statistical average. Numerical results show that similar theories not doing this make inaccurate predictions for the total density profile and the mean-squared displacement, which are observable quantities, in single-file diffusion experiments. Consequently, order-preserving dynamics provides a strong inductive argument for haecceitism. In classical statistical mechanics, it is therefore possible to distinguish between indistinguishable particles in the modal universe - via their names.

\section*{Acknowledgements}
I am very grateful to Paul M. N\"ager for his continuous support, and for detailed feedback and suggestions that helped me in developing this article. Moreover, I thank Ren{\'e} Wittmann for many helpful discussions about the physics of OPD. I also wish to thank Ulrich Krohs, Oliver Robert Scholz, Ansgar Seide, and the other participants of the thesis colloquium for helpful feedback on an earlier version of this work. In addition, I thank Raphael Wittkowski for introducing me to DDFT. Finally, I thank two anonymous referees for helpful comments. This work is funded by the Deutsche Forschungsgemeinschaft (DFG, German Research Foundation) -- grant number WI 4170/3-1. I also thank the Studienstiftung des deutschen Volkes for financial support.

\begin{flushright}
 \emph{Institut f\"ur Theoretische Physik\\
  Center for Soft Nanoscience\\
  Philosophisches Seminar\\
  Westf\"alische Wilhelms-Universit\"at M\"unster\\
  D-48149 M\"unster, Germany\\
  michael.tevrugt@uni-muenster.de
}
\end{flushright}


\end{document}